\documentclass[useAMS,usenatbib]{mn2e}
\usepackage{multirow}
\usepackage{epsfig}
\usepackage{subfigure}
\usepackage{url}

\makeatletter
\let\jnl@style=\rm
\def\ref@jnl#1{{\jnl@style#1}}

\def\aj{\ref@jnl{AJ}}                   
\def\araa{\ref@jnl{ARA\&A}}             
\def\apj{\ref@jnl{ApJ}}                 
\def\apjl{\ref@jnl{ApJ}}                
\def\apjs{\ref@jnl{ApJS}}               
\def\ao{\ref@jnl{Appl.~Opt.}}           
\def\apss{\ref@jnl{Ap\&SS}}             
\def\aap{\ref@jnl{A\&A}}                
\def\aapr{\ref@jnl{A\&A~Rev.}}          
\def\aaps{\ref@jnl{A\&AS}}              
\def\azh{\ref@jnl{AZh}}                 
\def\baas{\ref@jnl{BAAS}}               
\def\jrasc{\ref@jnl{JRASC}}             
\def\memras{\ref@jnl{MmRAS}}            
\def\mnras{\ref@jnl{MNRAS}}             
\def\pra{\ref@jnl{Phys.~Rev.~A}}        
\def\prb{\ref@jnl{Phys.~Rev.~B}}        
\def\prc{\ref@jnl{Phys.~Rev.~C}}        
\def\prd{\ref@jnl{Phys.~Rev.~D}}        
\def\pre{\ref@jnl{Phys.~Rev.~E}}        
\def\prl{\ref@jnl{Phys.~Rev.~Lett.}}    
\def\pasp{\ref@jnl{PASP}}               
\def\pasj{\ref@jnl{PASJ}}               
\def\qjras{\ref@jnl{QJRAS}}             
\def\skytel{\ref@jnl{S\&T}}             
\def\solphys{\ref@jnl{Sol.~Phys.}}      
\def\sovast{\ref@jnl{Soviet~Ast.}}      
\def\ssr{\ref@jnl{Space~Sci.~Rev.}}     
\def\zap{\ref@jnl{ZAp}}                 
\def\nat{\ref@jnl{Nature}}              
\def\iaucirc{\ref@jnl{IAU~Circ.}}       
\def\aplett{\ref@jnl{Astrophys.~Lett.}} 
\def\apspr{\ref@jnl{Astrophys.~Space~Phys.~Res.}}
\def\bain{\ref@jnl{Bull.~Astron.~Inst.~Netherlands}}
\def\fcp{\ref@jnl{Fund.~Cosmic~Phys.}}  
\def\gca{\ref@jnl{Geochim.~Cosmochim.~Acta}}   
\def\grl{\ref@jnl{Geophys.~Res.~Lett.}} 
\def\jcp{\ref@jnl{J.~Chem.~Phys.}}      
\def\jgr{\ref@jnl{J.~Geophys.~Res.}}    
\def\jqsrt{\ref@jnl{J.~Quant.~Spec.~Radiat.~Transf.}}
\def\memsai{\ref@jnl{Mem.~Soc.~Astron.~Italiana}}
\def\nphysa{\ref@jnl{Nucl.~Phys.~A}}   
\def\physrep{\ref@jnl{Phys.~Rep.}}   
\def\physscr{\ref@jnl{Phys.~Scr}}   
\def\planss{\ref@jnl{Planet.~Space~Sci.}}   
\def\procspie{\ref@jnl{Proc.~SPIE}}   

\makeatother

\newcommand {\apgt} {\ {\raise-.5ex\hbox{$\buildrel>\over\sim$}}\ }
\newcommand {\aplt} {\ {\raise-.5ex\hbox{$\buildrel<\over\sim$}}\ } 

\title[The hard X-ray emission of NGC 2110 observed by NuSTAR]{The Seyfert 2 galaxy NGC 2110: hard X-ray emission observed by {\it NuSTAR} and variability of the iron K$\alpha$ line}

\author[Andrea Marinucci, et al.]{A. Marinucci$^{1}$\thanks{E-mail: marinucci@fis.uniroma3.it (AM)}, G. Matt$^{1}$, S. Bianchi$^{1}$, T. N. Lu$^{2}$, P. Arevalo$^{3,4}$, M. Balokovi\'{c}$^{2}$,
\newauthor
D. Ballantyne$^{5}$, F. E. Bauer$^{3,6}$, S. E. Boggs$^{7}$,F. E. Christensen$^{8}$, W. W. Craig$^{8,9}$, 
\newauthor
P. Gandhi$^{10}$,C. J. Hailey$^{11}$,  F. Harrison$^{2}$, S. Puccetti$^{12,13}$, E. Rivers$^{2}$, 
\newauthor
 D. J. Walton$^{2}$, D. Stern$^{14}$ and  W. Zhang$^{15}$\\
$^1$Dipartimento di Matematica e Fisica, Universit\`a degli Studi Roma Tre, via della Vasca Navale 84, 00146 Roma, Italy\\
$^2$Cahill Center for Astronomy and Astrophysics, California Institute of Technology, Pasadena, CA, 91125 USA\\
$^3$Instituto de Astrof\'{i}sica, Facultad de F\'{i}sica, Pontificia Universidad Cat\'{o}lica de Chile, 306, Santiago 22, Chile\\
$^4$Instituto de F\'{i}sica y Astronom\'{i}a, Universidad de Valpara\'{i}so, Chile\\
$^5$Center for Relativistic Astrophysics, School of Physics, Georgia Institute of Technology, Atlanta, GA 30332, USA\\
$^6$Space Science Institute, 4750 Walnut Street, Suite 205, Boulder, Colorado 80301\\
$^7$Space Science Laboratory, University of California, Berkeley, California 94720, USA\\
$^8$DTU Space National Space Institute, Technical University of Denmark, Elektrovej 327, 2800 Lyngby, Denmark\\
$^9$Lawrence Livermore National Laboratory, Livermore, California 94550, USA\\
$^{10}$Department of Physics, University of Durham, South Road, Durham, DH1 3LE, UK\\
$^{11}$Columbia Astrophysics Laboratory, Columbia University, New York, New York 10027, US\\
$^{12}$ASDC-ASI, Via del Politecnico, 00133 Roma, Italy\\
$^{13}$INAF Osservatorio Astronomico di Roma, via Frascati 33,00040 Monte Porzio Catone (RM), Italy\\
$^{14}$Jet Propulsion Laboratory, California Institute of Technology, 4800 Oak Grove Drive, Pasadena, CA 91109, USA\\
$^{15}$NASA Goddard Space Flight Center, Greenbelt, Maryland 20771, USA\\
}

\begin{document}
\maketitle
\label{firstpage}

\begin{abstract} 
 We present {\it NuSTAR} observations of the bright Seyfert 2 galaxy NGC 2110 obtained in 2012, when the source was at the highest flux level ever observed, and in 2013, when the source was at a more typical flux level. We include archival observations from other X-ray satellites, namely {\it XMM-Newton}, {\it Suzaku}, {\it BeppoSAX}, {\it Chandra} and {\it Swift}. Simultaneous {\it NuSTAR} and {\it Swift} broad band spectra (in the 3-80 keV range) indicate a cutoff energy $E_{\rm c}>210$ keV, with no detectable contribution from Compton reflection. NGC 2110 is one of the very few sources where no evidence for distant Compton thick scattering is found and, by using temporal information collected over more than a decade, we investigate variations of the iron K$\alpha$ line on time scales of years. The Fe K$\alpha$ line is likely the sum of two components: one constant (originating from distant Compton-thick material) and the other one variable and linearly correlated with the source flux (possibly arising from Compton-thin material much closer to the black hole).
\end{abstract}


\section{ Introduction}
The X-ray spectra of Seyfert 2 galaxies offer a unique opportunity to probe the circumnuclear environment of Active Galactic Nuclei (AGN). The obscuration of the primary radiation by matter with column densities typically in the $10^{22}-10^{24}$ cm$^{-2}$ range allow a study of the cold and ionised reflectors that cannot be observed, due to dilution effects, in unobscured sources. Typical X-ray features of the cold circumnuclear material include an intense Fe K$\alpha$ line at 6.4~keV due to fluorescence emission and a reflection continuum peaking at $\sim30$ keV \citep{mpp91,gf91}.

The primary continuum is thought to arise from the innermost regions surrounding the central black hole, in a hot corona above the accretion disc. It is observed as a power-law spectrum extending to energies determined by the electron temperature in the hot corona \citep{rl79}. The power-law index is a function of the plasma temperature $T$ and optical depth $\tau$. 

NGC 2110, at a redshift $z=0.008$, is one of the brightest Seyfert galaxies in the hard X-ray ($>$10 keV) band and it shows a prominent Fe K$\alpha$ line accompanied by variable intrinsic emission \citep{mush82, hka96}. It has been intensively studied by most X-ray observatories, and it has shown very interesting and peculiar characteristics. \citet{mbc99} reported a photon index of $\Gamma=1.86$ analysing {\it BeppoSAX} PDS data above 13 keV \citep[consistent with typical values found in Seyfert 1 sources:][]{np94}. However, when 2-10 keV data are considered, the photon index becomes flatter ( $\Gamma=1.67$) and the authors suggested the presence of obscuring material with a column density $N_{\rm H}=4.1^{+0.5}_{-0.3}\times 10^{22}$ cm$^{-2}$ along the line of sight and an additional absorber with $N_{\rm H}\sim3 \times 10^{23}$ cm$^{-2}$, partially covering the nuclear source. This scenario was confirmed by {\it XMM-Newton}, {\it Chandra} and {\it Suzaku} observations \citep{elt07,rfk06,rmr14}. 

NGC 2110 is one of the very few Seyfert galaxies that, despite the intense iron K$\alpha$ emission line at 6.4 keV, does not show any Compton reflection from circumnuclear material: values of $R \leq0.17$ and $R\leq0.1$ were found with {\it BeppoSAX} and {\it Suzaku}, respectively \citep{mbc99,rmr14}. If the line emitting material is Compton-thick ($N_{\rm H}>10^{24}$ cm$^{-2}$), the iron K$\alpha$ emission would be accompanied by a Compton reflection component above 10 keV \citep{mpp91,gf91}. The inferred upper limits on the reflection fraction $R$ in this source suggest that the iron K$\alpha$ line is not produced by distant, Compton-thick material but is instead emitted by Compton-thin matter, such as in the case of NGC 7213 \citep{bianchi03b, bianchi08b}.

Recently, {\it NuSTAR} observed NGC 2110 in an extremely bright state. The lack of reflection components in NGC 2110 arising from the accretion disc or from the putative torus above 10 keV makes this source a perfect candidate for measuring a high-energy cutoff. We present a detailed multi-epoch X-ray study of NGC 2110 including two recent {\it NuSTAR} observations, obtained in 2012 and 2013, with the dual aims of studying the behaviour of the Fe K$\alpha$ line with respect to the highly variable intrinsic continuum and of constraining the high energy cutoff in this source.

\section{Observations and data reduction}
\subsection{NuSTAR}
{\it NuSTAR} (Harrison et al. 2013) observed NGC 2110 with its two coaligned X-ray telescopes with corresponding focal planes: Focal Plane Module A (FPMA) and B (FPMB) starting on 2012 October 5 and 2013 February 14 for a total of $\sim32$ ks and $\sim26$ ks of elapsed time, respectively.  The Level 1 data products were processed with the {\it NuSTAR} Data Analysis Software (NuSTARDAS) package (v. 1.3.0). Cleaned event files (level 2 data products) were produced and calibrated using standard filtering criteria with the \textsc{nupipeline} task and the latest calibration files available in the {\it NuSTAR} calibration database (CALDB). Both extraction radii for the source and background spectra were $1.5$ arcmin. After this process, the net exposure times for the two observations were 15 ks and 12 ks, with 3-80 keV count rates of $6.53\pm0.02$ and $4.50\pm 0.02$ cts s$^{-1}$ for FPMA, and $6.32\pm0.02$ and $4.25\pm 0.02$ cts s$^{-1}$ for FPMB. 
The two pairs of {\it NuSTAR} spectra were binned in order to over-sample the instrumental resolution by at least a factor of 2.5 and to have a Signal-to-Noise Ratio (SNR) greater than 5$\sigma$ in each spectral channel.

\subsection{Suzaku}
NGC 2110 was observed by {\it Suzaku} starting on 2005 September 16 (OBSID 100024010) and seven years later, starting on 2012 August 31 (OBSID 707034010). Data were taken from the X-ray Imaging Spectrometer (XIS) and the Hard X-ray Detector (HXD). We reprocessed the event files with the latest calibration files available (2014-02-03) using \textsc{ftools} 6.14 and \textsc{suzaku} software Version 21, adopting standard procedures and recommended screening criteria. The source extraction radius was 1.8 arcmin. Background spectra were extracted from source-free regions of 1.8 arcmin radius. Response matrices and ancillary response files were generated using \textsc{xisrmfgen} and \textsc{xisarfgen}.
The 0.5-10 keV spectra extracted from the front-illuminated XIS0 and XIS3 were co-added using the ftool \textsc{addascaspec}, for net exposure times of 102 ks and 103 ks for the two data sets. Spectra were binned in order to over-sample the instrumental resolution by at least a factor of three and to have no less than 30 counts in each background-subtracted spectral channel. This allows the applicability of $\chi^2$ statistics.

We reduced the HXD PIN data  using the \textsc{aepipeline} reprocessing tool, and for background determination we downloaded and utilized the tuned non-X-ray background (NXB) provided by the HXD team. We extracted source and background spectra using the same good time intervals.

\subsection{XMM-Newton}
NGC 2110 was observed by {\it XMM-Newton} \citep{xmm} for $\sim$60 ks, starting on 2003 March 5 (OBSID 0145670101) with the EPIC CCD cameras, the Pn \citep{struder01} and the two MOS \citep{turner01}, operated in large window and thin filter mode. Data from the MOS detectors are not included in our analysis due to the lower statistics of the spectra. The extraction radii and the optimal time cuts for flaring particle background were computed with SAS 13 \citep{gabr04} via an iterative process which leads to a maximization of the SNR, similar to the approach described in \citet{pico04}. The resulting optimal extraction radius was 38 arcsec and the background spectra were extracted from source-free circular regions with a radius of about 50 arcsec. 

After this process, the net exposure time was 44 ks for the EPIC-Pn. Spectra were binned in order to over-sample the instrumental resolution by at least a factor of three and to have no less than 30 counts in each background-subtracted spectral channel.

\subsection{Swift}
There were 9 {\it Swift}/XRT observations of NGC 2110, in 2006, 2008, 2009, 2012 and 2013. The first 8 observations were carried out with the photon counting (PC) mode, while the last observation was in window timing (WT) mode (Table \ref{swiftobs}). ObsIDs 80364001 and 80364002 are simultaneous with {\it NuSTAR}. We reprocessed all the datasets to generate cleaned event files with the \textsc{xrtpipeline} script included in the \textsc{heasoft} version 6.13. The PC mode
observations are all affected by pile-up, while the final observation, obtained in WT mode, does not have pile-up issues. To correct the pile-up, we extracted the source spectrum within a 72 arcsec radius circular region, excluding the central 10 arcsec radius aperture for the PC mode datasets. Obs. IDs, dates and net exposure times are listed in Table \ref{swiftobs}.

\subsection{Chandra}
NGC 2110 was observed by {\it Chandra} on 2000-04-22 with the ACIS-S camera (Obs. ID 883), on 2001-12-19 with three consecutive HETGs pointings (Obs. ID 3143, 3417 and 3418) and on 2003-03-05 with the HETGs (Obs. ID 4377). All these observations are discussed in detail in \citet{elt07} where the four HETG spectra were merged. Since pile-up was found in the ACIS-S observation, we do not use this spectrum in our analysis. Data were reduced using \textsc{ciao} 4.5 and the latest CALDB files, we merged Obs. IDs 3143, 3417 and 3418 spectra using the \textsc{add$_-$grating$_-$order} and \textsc{add$_-$grating$_-$spectra} tools. The resulting exposure times are 105 ks and 95 ks for the 2001 and 2003 HETGs data, respectively. Data were binned to have no less than 30 counts in each spectral channel.

\subsection{BeppoSAX}
The source was observed by {\it BeppoSAX} on 1997-10-12 with the MECS for a net exposure time of 76 ks. Reduced data were downloaded from the {\it BeppoSAX} online data archive \footnote[1]{available at:\\ {\tt http://www.asdc.asi.it/mmia/index.php?mission=saxnfi}}.

\begin{table}
\begin{center}
\hspace{0.3cm}
\begin{tabular}{c|ccc}
Obs. ID & Date & $\rm T_{exp}$ (ks) \\
 \hline
 \hline
 35459001 &2006-03-25 &8.3 \\
35459002 &2006-04-08 & 9.2\\
35459003 &2006-04-15 & 2.3\\
35459004 &2008-08-31 & 2.2\\
35459005 &2009-10-12 & 3.5\\
80364001* &2012-10-05 & 7.1\\
35459006 &2013-02-03 & 14.1\\
80364002* &2013-02-15 & 0.9\\
35459007 &2013-03-09 & 2.7\\
\hline
\hline
\end{tabular}\\
\end{center}
\caption{\label{swiftobs} Observation log for the {\it Swift} monitoring of NGC 2110. Observations IDs, dates and net exposure times (after filtering and correction for photon pile-up) are reported. Asterisks indicate observations simultaneous with {\it NuSTAR}.}
\end{table}

\section{Spectral analysis}
We first study the simultaneous {\it NuSTAR} and {\it Swift} data to probe the primary radiation from the AGN and the properties of the hot corona. We then perform a multi-epoch phenomenological X-ray analysis to study the behavior of the Fe K$\alpha$ emission line in response to variability of the nuclear continuum. 

In previous work, broad band analyses of some data sets revealed the presence of extra-nuclear emission in the softer (E$<$1 keV) part of the spectra \citep{elk06,rfk06,elt07,rmr14}, the analysis of this component is beyond the scope of this work. We focus our analysis on the 3-79 keV band where the contribution from soft diffuse emission is negligible.

The adopted cosmological parameters are $H_0=70$ km s$^{-1}$ Mpc$^{-1}$, $\Omega_\Lambda=0.73$ and $\Omega_m=0.27$, i.e. the default ones in \textsc{xspec 12.8.1} \citep{xspec}. Errors correspond to the 90\% confidence level for one interesting parameter ($\Delta\chi^2=2.7$), if not stated otherwise. 

\subsection{\label{380analysis}{\it NuSTAR} 3--80 keV spectral analysis}
\begin{figure} 
\begin{center}
\epsfig{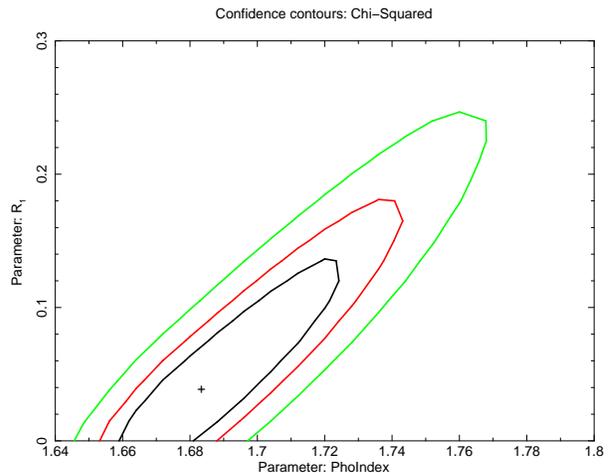}
\caption{\label{R} Contour plot between  reflection fraction $R$ and photon index $\Gamma$ for the high flux {\it NuSTAR} observation in 2012, when only 10--80 keV data are considered. The solid black, red and green lines correspond to 68$\%$, 90$\%$ and 99$\%$ confidence levels, respectively.}
\end{center}
\end{figure}
\begin{figure} 
\begin{center}
\epsfig{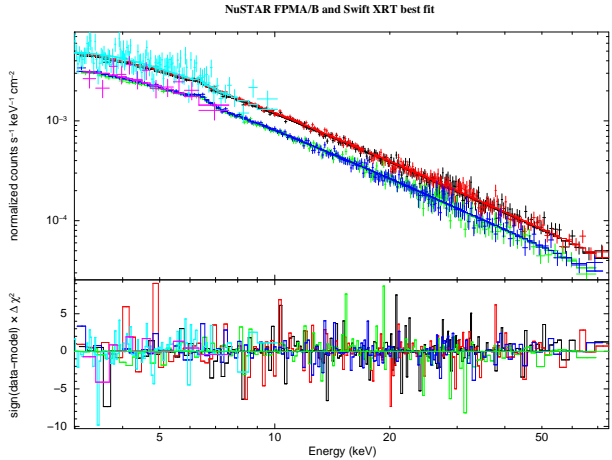}
\caption{\label{nufit}  3-80 keV simultaneous {\it NuSTAR}+{\it Swift} best fit. {\it NuSTAR} FPMA and FPMB spectra are in black and red for the 2012 observation, green and blue for the 2013 one. {\it Swift} XRT 2012 data are in light blue and 2013 spectra in magenta. Residuals are shown in the bottom panel, when a model composed of an absorbed cutoff power law and a Gaussian line at 6.4 keV is applied to our data set. No strong residuals are present accross the whole energy band.}
\end{center}
\end{figure}
\begin{figure}
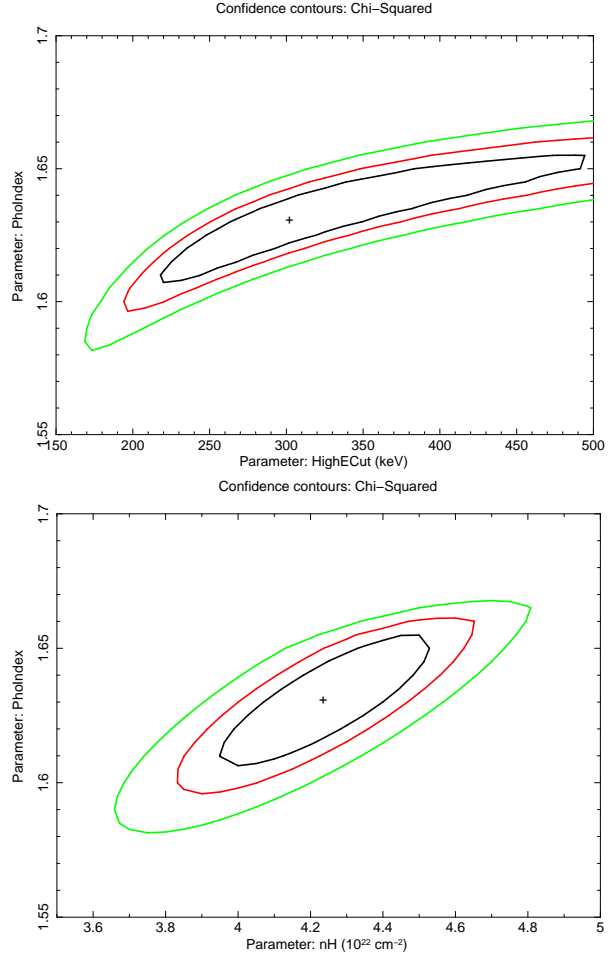

\begin{center}
\epsfig{file=cont_ec.ps, angle=-90, width=\columnwidth}
\epsfig{file=cont_nh_gamma.ps, angle=-90, width=\columnwidth}
\caption{\label{ec} {\it Top panel:} Contour plot between photon index $\Gamma$ and cutoff energy $E\rm _c$, when simultaneous 3--80 keV {\it Swift}+{\it NuSTAR} data are considered. The solid black, red and green lines correspond to 68$\%$, 90$\%$ and 99$\%$ confidence levels, respectively. High-energy cutoff values are limited to $\leq$500 keV because it is the maximum value allowed in the \textsc{cutoffpl} model.  {\it Bottom panel:} Contour plot between photon index $\Gamma$ and absorbing column density along the line of sight.}
\end{center}
\end{figure}

The X-ray spectra of NGC 2110 have been widely studied in the past few years and several analyses have shown the presence of a $\Gamma\sim1.7$ power law continuum partially covered by several layers of absorbing material with column densities in the range 2--6$\times 10^{22}$ cm$^{-2}$ \citep{elt07,rmr14}. Above 10 keV the effect of this absorbing material can be neglected and we therefore start our spectral analysis fitting the {\it NuSTAR} spectra in the 10--80 keV range, to have a direct measurement of the photon index of the primary power law. We will test \textit{a posteriori} if this assumption is correct.

Our model is composed of a power law, with the only free parameters in the fit being the photon index, the normalization of the power law and the cross-calibration factors between the FPMs. When the model is applied to the {\it NuSTAR} 2012 and 2013 spectra, the fit is good ($\chi^2$/dof=568/601=0.95) and no strong residuals are present across the whole energy band. The best fit photon index is $\Gamma=1.691\pm0.015$ and the cross-calibration factors are $\rm K_{A-B}^{2012}=1.017\pm0.015$ and $\rm K_{A-B}^{2013}=1.000\pm0.020$. If we leave the photon indices of the 2012 and 2013 observations free to vary independently no improvement in the fit is found and they are both consistent with the inferred single best fit value. A neutral absorber along the line of sight does not improve the fit ($\Delta \chi^2=2$ for one additional variable parameter) and only an upper limit $N_{\rm H}<8\times 10^{22}$ cm$^{-2}$ is found: this confirms that the circumnuclear absorbers in this object do not affect the analysis of the 10-80 keV {\it NuSTAR} spectra.

\begin{table*}
\begin{center}
\hspace{0.3cm}
\begin{tabular}{c|ccccccccc}
Instrument & Date & $N_{\rm H}$& $\Gamma$ & Fe K$\alpha$ En.& $\sigma$&EW& $F_{\rm K\alpha}$& $F_{\rm K\beta}$& $F_{\rm 3-10\ keV}$\\
 \hline
 \hline
   {\it BeppoSAX }& 1997-10-12&$4.3\pm0.9$& $1.74\pm0.09$&$6.43^{+0.06}_{-0.09}$&$<280$& $194^{+69}_{-50}$&$8.3^{+3.0}_{-2.3}$&$<1.3$&$2.77\pm0.05$\\
 {\it Chandra} & 2001-12-19&$4.0\pm1.8$& $1.67^{+0.30}_{-0.25}$&$6.400\pm0.008$&$16^{+14}_{-10}$& $90^{+30}_{-25}$&$5.4^{+1.8}_{-1.5}$&$<2.8$&$3.84\pm0.07$\\
  {\it Chandra }& 2003-03-05&$<4.5$& $1.25^{+0.48}_{-0.33}$&$6.391\pm0.016$&$30^{+31}_{-16}$& $135^{+60}_{-45}$&$5.5^{+2.5}_{-2.0}$&$<2.0$&$2.80\pm0.07$\\
  {\it XMM} & 2003-03-05&$3.9\pm0.4$& $1.57\pm0.05$&$6.42\pm0.01$&$62\pm14$& $145\pm15$&$5.0\pm0.5$&$0.8\pm0.3$&$2.26\pm0.03$\\
  {\it Suzaku} & 2005-09-16& $3.8\pm0.2$&$1.63\pm0.02$&$6.40\pm0.01$&$50\pm15$& $55\pm5$&$8.4\pm0.8$&$0.6\pm0.5$&$9.90\pm0.03$\\
 {\it Suzaku}& 2012-08-31& $4.5\pm0.2$&$1.63\pm0.02$&$6.39\pm0.01$&$<55$& $50\pm7$&$9.7\pm0.9$&$<1.0$&$11.8\pm0.1$\\
{\it NuSTAR}& 2012-10-05&$4.0\pm0.4$& $1.64\pm0.03$&$6.33\pm0.07$&$<192$& $35\pm10$&$9.5\pm3.0$&$<2.3$&$17.1\pm0.2$\\
 {\it NuSTAR} & 2013-02-14&$4.0\pm0.7$&$1.64\pm0.05$&$6.45\pm0.07$&$175^{+200}_{-140}$& $90^{+42}_{-25}$&$16.0^{+8.0}_{-4.0}$&$<3.4$&$11.7\pm0.2$\\
\hline
\hline
 & & & & & & & & & \\
 \multicolumn{10}{c}{\large{{\it Swift} best fit parameters}}\\
 & & & & & & & & & \\
 $\chi^2$/d.o.f. & Date & $N_{\rm H}$& $\Gamma$ &  Fe K$\alpha$ En.& $\sigma$&EW& $F_{\rm K\alpha}$& $F_{\rm K\beta}$& $F_{\rm 3-10\ keV}$\\
 \hline
 \hline
  59/61 & 2006-03-25& $3.5\pm2.5$& $1.35\pm0.45$&$6.50\pm0.15$&$115^{+80}_{-85}$& $190\pm115$&$25.0\pm15.0$&$<13$&$8.9\pm0.5$\\
  83/85&2006-04-08 &  $3.0\pm2.0$&$1.25\pm0.35$&$6.50^{+0.11}_{-0.08}$&$<250$&$100^{+80}_{-64}$ &$14.7^{+10.2}_{-9.3}$&$<7$&$9.8\pm0.5$\\
  15/24 &2006-04-15& $4.0\pm2.3$&$1.7$*&$6.7\pm0.5$&$<350$& $450^{+665}_{-400}$&$50.9^{+78.5}_{-44.5}$&$<34$&$8.1\pm0.9$\\
 17/15 &2008-08-31 &$7.5\pm3.0$& $1.7$*&$6.4$*&$60$*& $<95$&$<9.0$&$<44$&$5.9\pm0.9$\\
  37/35&2009-10-12 & $5.5\pm3.5$ & $1.43^{+0.70}_{-0.51}$&$6.2\pm0.1$ &$<350$ & $170^{+120}_{-120}$ & $35\pm25$& $<38$& $11.4\pm0.8$\\
  106/94 &2012-10-05 & $5.9\pm2.5$& $1.7^{+0.3}_{-0.3}$&$6.4$* & $60$*& $<35$ &$<30.7$ &$<15$ & $17.3\pm0.4$\\
  170/159 &2013-02-03 & $6.3\pm1.9$& $1.75^{+0.25}_{-0.25}$& $6.4$*& $60$*& $<50$ & $<14$&$<13$ & $16.0\pm0.9$\\
  9/8 &2013-02-15 & $7.0\pm3.5$&$1.7$* & $6.4$*& $60$*& $<250$ &$<53$ &$<30$ &$12.5\pm1.6$ \\
  81/56&2013-03-09 &  $4.3\pm1.2$& $1.7$* &$6.2\pm0.1$& $60$*& $160\pm100$&$22.7\pm12.5$  &$<10$ &$8.0\pm0.8$ \\
\hline
\hline
\end{tabular}\\
\end{center}
\caption{\label{refl_best_fit} Best fit parameters when the 3-10 keV phenomenological fit is applied to the data. Column densities are in $10^{22}$ cm$^{-2}$ units, energy centroids are in keV units, equivalent widths (EW) and widths ($\sigma$) of the Fe K$\alpha$ line in eV units, line fluxes are in $10^{-5}$ ph cm$^{-2}$ s$^{-1}$ units and 3-10 keV observed fluxes are in 10$^{-11}$ erg cm$^{-2}$ s$^{-1}$ units.}
\end{table*}

In the past, high energy observations of NGC 2110 only revealed upper limits for the fraction $R$ of the Compton reflection of the primary continuum by cold, neutral material \citep{mbc99,rmr14} which is usually found in Seyfert galaxies \citep{dad08,rwc11}. The addition of such a component in our fit, modeled with \textsc{pexrav} \citep{mz95} with $\Gamma$ fixed to that of the primary continuum and the inclination angle to 60 degrees, does not lead to an improvement of the fit ($\Delta \chi^2=2$ for two additional variable parameters) and only upper limits $R_{2012}<0.15$ and $R_{2013}<0.14$ in the reflection components are found for the 2012 and 2013 observations, respectively. The contour plot between $R_{2012}$ and $\Gamma$ is shown in Fig. \ref{R}. Other parameters do not differ from best the fit values presented above. When we add a cutoff power law in our fit (model \textsc{cutoffpl} in \textsc{xspec}) no significant improvement is found ($\Delta \chi^2=2$ for one additional variable parameter): we find a best fit value of $\Gamma=1.647_{-0.053}^{+0.014}$ and a lower limit $E\rm _c>230$ keV for the cutoff energy.

\subsection{\label{380analysis_swift}{\it NuSTAR}$+${\it Swift} 3--80 keV spectral analysis}

We then consider 2012 and 2013 {\it NuSTAR} data down to 3 keV and introduce the simultaneous {\it Swift}-XRT spectra, with net exposure times of 7.1 ks and 0.9 ks, respectively. 

The model is composed of an absorbed power law with a high-energy cutoff and two Gaussian lines, to reproduce the Fe K$\alpha$ and K$\beta$ emission at 6.4 keV and 7.056 keV, respectively. We then convolved the baseline model with a Galactic column density $N_{\rm H}^{\rm G}=1.62\times 10^{21}$ cm$^{-2}$ \citep{kalberla05}, modeled with \textsc{tbabs} in \textsc{xspec}, using solar abundances \citep{wilms00} and cross-sections from \citet{verner96}. In \textsc{xspec} the model reads as follows: \textsc{tbabs*zwabs*(cutoffpl+zgauss+zgauss)}. XRT-FPMA cross-calibration factors are introduced as a variable parameters. The fit is good ($\chi^2$/dof=876/881=0.99) and no strong residuals are seen (Fig. \ref{nufit}). 

The inclusion of Compton reflection in the fit leads to an insignificant improvement of the fit ($\Delta \chi^2=2$ for two additional variable parameters) and its contribution to the total 3--80 keV flux is $F^{2012}_{\rm refl}<1.5\%$ and $F^{2013}_{\rm refl}<2.5\%$. Best fit cross-calibration factors are $K\rm_{XRT-FPMA}^{2012}=0.98\pm0.03$ and $K\rm_{XRT-FPMA}^{2013}=0.95\pm0.09$. We measure a lower limit for the high energy cutoff $E\rm _c>210$ keV and in Fig. \ref{ec} the contour plots of E$\rm _c$ vs $\Gamma$ as well as $N_{\rm H}$ vs $\Gamma$ are shown.  Best fit values do not significantly differ from the ones discussed in Sect. \ref{380analysis}.

We next use a more physical model \citep[\textsc{compTT} in \textsc{xspec};][]{tit94} to measure the coronal temperature $kT_e$ and optical depth $\tau$. In this model the soft seed photon spectrum is a Wien law; we fixed the disc temperature to 30 eV, appropriate for $M_{\rm BH}\approx 10^8$ M$_\odot$ \citep[the black hole mass in NGC 2110 is estimated to be $M_{\rm BH}\sim2\times 10^8$ M$_\odot$ via the $M_{\rm BH}$--$\sigma$ relation;][]{mbe07}. In the case of a slab geometry the fit is good ($\chi^2$/dof$=873/880=0.99$) and best fit parameters $kT_e=190\pm130$ keV and $\tau=0.22^{+0.51}_{-0.05}$ are found. 

The measured iron K$\alpha$ equivalent width for the 2012 and 2013 observations (EW$\simeq$35--200 eV, Table \ref{refl_best_fit}) are unusually large given the observed Compton reflection for the line to originate from reflection from Compton thick material. We therefore confirm the result in \citet{mbc99} from the low flux {\it BeppoSAX} observation and with {\it Suzaku} \citep{rfk06, rmr14}.
If the Compton reflection and the iron K$\alpha$ emission line arise from the same distant, Compton-thick material, they can be self-consistently modelled using the \textsc{pexmon} model in \textsc{Xspec} \citep{nan07}, with the inclination angle fixed at $60$ degrees. This leads to a best fit $\chi^2$/dof=930/884=1.06 with visible residuals around iron K$\alpha$. Leaving the iron abundance free to vary a significant improvement in the fit is found ($\Delta \chi^2=34$ for one additional variable parameter) and we find a $A_{\rm Fe}>22$ with respect to the solar value. No significant variation between photon index, reflection fraction, absorbing column densities, cross-calibration factors and best fit parameters discussed above is found. Such a large iron abundance is unrealistic and therefore we discard this scenario. 

The alternative scenario we consider is the one in which the line is produced by Compton-thin material ($N_{\rm H}= 10^{22}-10^{23}$ cm$^{-2}$) that does not contribute significantly to the Compton reflection \citep{mgm03}. We model the Compton reflection and iron fluorescent lines with \textsc{MYTorus} scattered and line components \citep{my09, yaq12}. We add a further component to reproduce absorbing material along the line of sight. The assumed geometry is a torus of gas and dust with a 60 degree opening angle.
When we apply this model to the 2012 and 2013 {\it Swift}+{\it NuSTAR} simultaneous data, we leave the normalization of the primary continua and the column density along the line of sight as variable parameters. Normalizations and column densities of the scattered and line components are linked and free to vary. The fit is good ($\chi^2$/dof$=907/884=1.02$) and we find no variations from best fit parameters discussed above are found. The best fit value for the column density of the scattering material producing the iron K$\alpha$ is $N_{\rm H}= 2.0\pm1.1\times10^{23}$ cm$^{-2}$. The ratios between the scattered and primary components' normalizations is consistent with unity, the standard \textsc{MyTorus} configuration \citep[coupled reprocessor model in][]{yaq12}, in which the torus is not aligned along the line of sight.\\
The value of the column density of the iron K$\alpha$ emitting gas is consistent with the estimate presented in \citet{bianchi03b} ($N_{\rm H}\simeq 3\times10^{23}$ cm$^{-2}$) for the case of NGC 7213, where the authors assumed a Broad Line Region covering factor $f_{\rm c}=0.35$, an EW$\simeq 100$ eV and a photon index $\Gamma=1.69$. This suggests that the two sources do indeed present similar features in the X-ray band.

\subsection{\label{specanalysis}Time history of the iron K$\alpha$ line}
\begin{figure}
\begin{center}
\epsfig{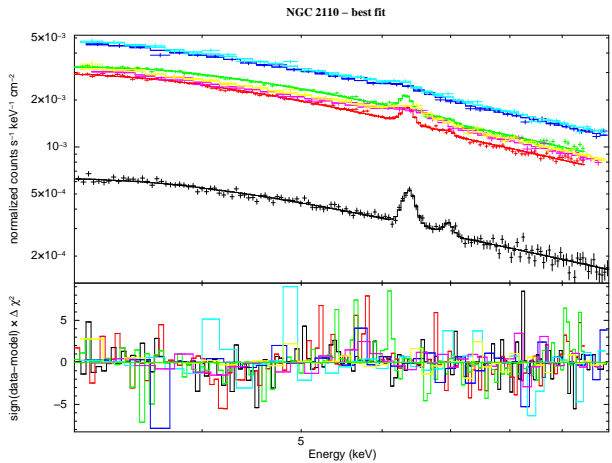}
\caption{\label{bestfit1} 3-10 keV phenomenological best fit. EPIC-Pn data are in black, {\it Suzaku} data from the 2005 and 2013 observations are in red and green, respectively. {\it NuSTAR} FPMA and FPMB data from the 2012 observation are in light and dark blue while {\it NuSTAR} FPMA and FPMB spectra obtained in 2013 are in orange and yellow, respectively. }
\end{center}
\end{figure}

We apply the model described in Sect. \ref{380analysis_swift} (an absorbed primary continuum and two Gaussian lines) to the archival 3-10 keV data sets. This phenomenological fit is intended for studying the Fe K$\alpha$ equivalent width and flux on time scales of months and years.  

We first fit data with high SNR from {\it XMM-Newton}, {\it Suzaku} and {\it NuSTAR} with the above model:
the fit is good ($\chi^2$/dof=628/544=1.15) and no strong residuals are present across the energy band (Fig. \ref{bestfit1}). The best fit parameters are reported in Table 2.  
If we introduce a reflection component in our fit, no variation in the parameters of the lines is found with respect to the best fit values.

Then, we fit the 3-10 keV spectra of the nine {\it Swift} snapshots of the source, fixing the photon index to $\Gamma=1.7$ (as inferred from the high energy data analysis).Further parameters such as centroids and widths of the Fe K$\alpha$ emission line are fixed to 6.4 keV and 60 eV (as inferred from the {\it XMM} and {\it Suzaku} analyses), respectively, where SNR was too poor to accurately measure these parameters. Best fit $\chi^2$/dof values are reported in Table 2.

\begin{figure} 
\begin{center}
\epsfig{file=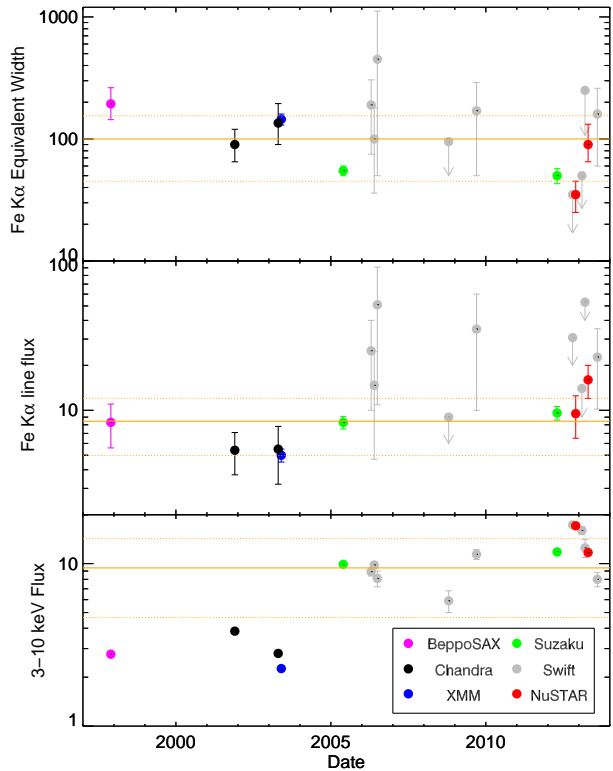, width=\columnwidth}
\caption{\label{lc} Time evolution of iron K$\alpha$ parameters and 3-10 keV observed flux of the source (not corrected for absorption). Equivalent widths are in eV units, line fluxes are in $10^{-5}$ ph cm$^{-2}$ s$^{-1}$ units and observed 3-10 keV fluxes in 10$^{-11}$ erg cm$^{-2}$ s$^{-1}$ units. Solid and dashed horizontal lines represent mean and standard deviations, respectively.}
\end{center}
\end{figure}

Fig. \ref{lc} shows the time evolution of the equivalent width of the Fe K$\alpha$ line flux and the observed 3-10 keV flux of the source. The object has a variable intrinsic emission in the 3-10 keV energy range: we measure a factor of $\sim$7.5 between the 2003 {\it XMM-Newton} and 2012 {\it NuSTAR} observations, while we find a factor $\sim1.5$ between the {\it Swift} observation on 2013-02-03 and the {\it NuSTAR} observation on 2013-02-14 on a 10 day time scale.

The variability of the Fe K$\alpha$ emission line carries information about the distance and ionization state of the emitting material (width and energy centroid). Indeed, the measured  Fe K$\alpha$ flux in the 2003 {\it XMM} observation is significantly lower than the one measured with {\it Suzaku} three years later. 

\begin{figure*} 
\begin{center}
\epsfig{file=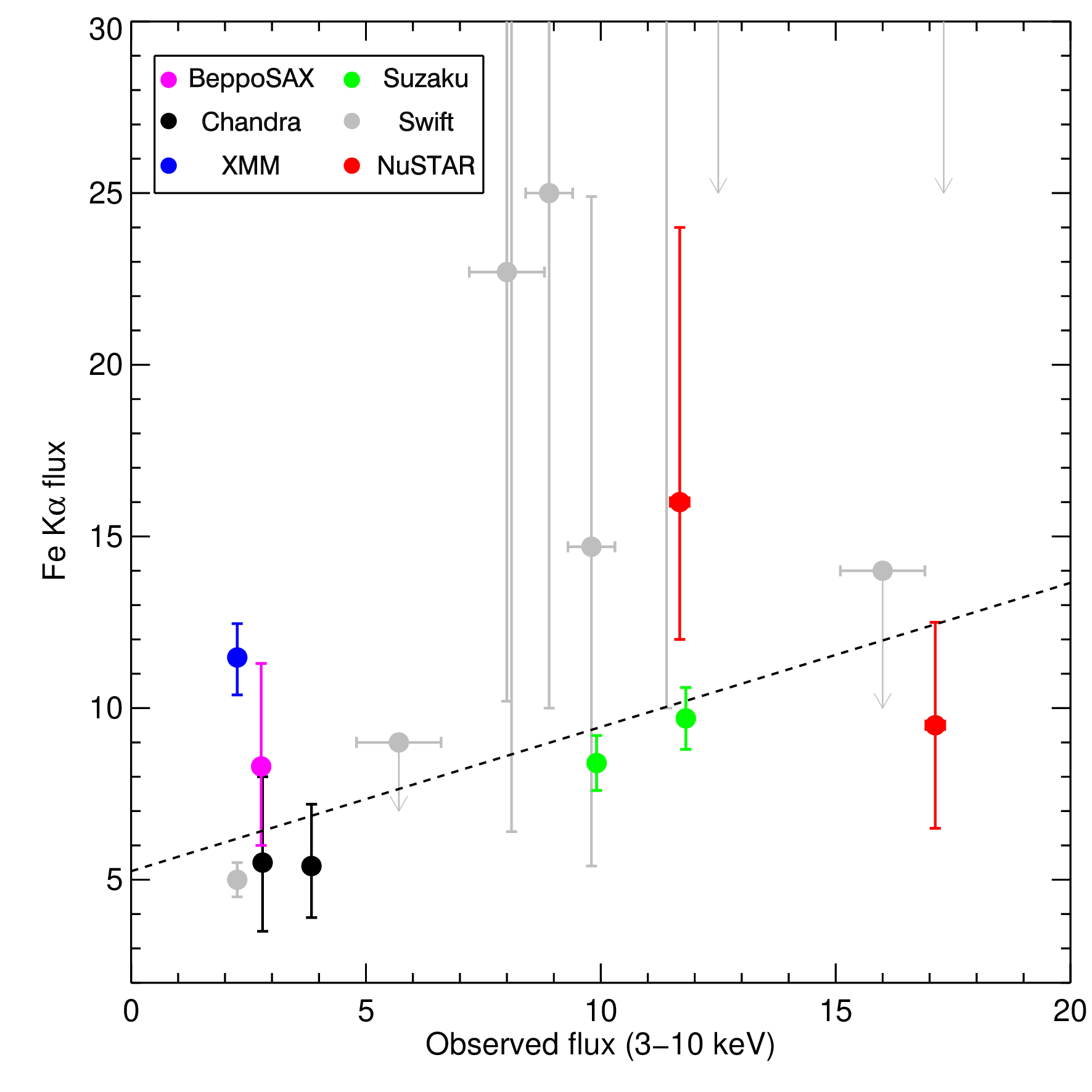, width=\columnwidth}
\epsfig{file=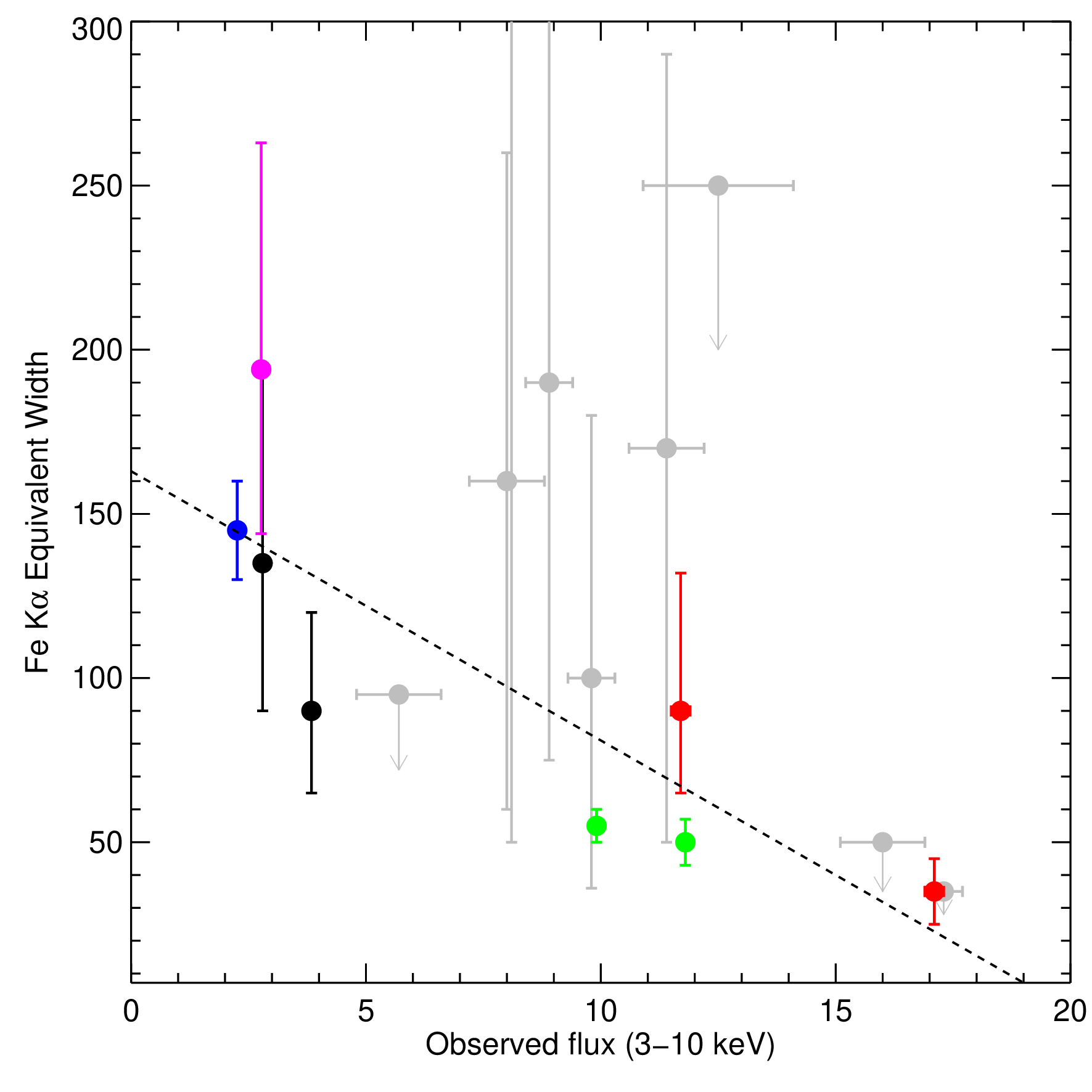, width=\columnwidth}
\caption{\label{lines} {\it Left panel:} Iron K$\alpha$ flux (in $10^{-5}$ ph cm$^{-2}$ s$^{-1}$ units) plotted against intrinsic 3-10 keV flux (in 10$^{-11}$ erg cm$^{-2}$ s$^{-1}$ units). {\it Right panel:} Iron K$\alpha$ equivalent widths (in eV units) plotted against intrinsic 3-10 keV flux (in 10$^{-11}$ erg cm$^{-2}$ s$^{-1}$ units). Dashed black lines represent best fit relations, when a linear fit is applied using only high SNR observations (i.e. removing {\it Swift} data points). }
\end{center}
\end{figure*}

In Fig. \ref{lines} we plot Fe K$\alpha$ fluxes and equivalent widths (EW) vs the observed 3-10 keV flux ($F\rm^c_{3-10}$), respectively. We examined the correlations between EW and $F\rm^c_{3-10}$ and between the Fe K$\alpha$ flux and $F\rm^c_{3-10}$ by performing a linear fit using only high SNR observations (i.e. removing {\it Swift} data points). 

We find a best fit relation ${\rm EW}=(-8.2\pm1.6)\times F{\rm^c_{3-10}}+(163\pm20)$ with a Spearman's rank correlation coefficient $\rho=-0.88\pm0.06$ and a null-hypothesis-probability of $5.5\times10^{-3}$. The other best fit relation is $F_{K\alpha}=(0.42\pm0.17)\times F{\rm^c_{3-10}}+(5.2\pm1.0)$ with a Spearman's rank correlation coefficient $\rho=0.67\pm0.09$ and a null-hypothesis-probability of $9.1\times10^{-2}$. Both best fit curves are shown in Fig. \ref{lines}. When we introduce the {\it Swift} data points in the fit, we obtain less statistically significant relations, with null-hypothesis-probabilities of $3.9\times10^{-2}$ and $0.37$ for the EW--$F^c_{3-10}$ and $F_{K\alpha}$--$F^c_{3-10}$ relations, respectively.

\section{Discussion}

\subsection{Properties of the hot corona}
Recently, the {\it NuSTAR}'s high sensitivity above 10 keV has led to high-energy cutoff measurements in a number of nearby Seyfert galaxies: IC 4329A \citep[$178^{+74}_{-40}$ keV,][]{bmf14}, SWIFT J2127.4+5654 \citep[$108^{+11}_{-10}$ keV,][]{mmk14}, Ark 120 \citep[$E_{\rm c}> 190$ keV,][]{mmg14} and 3C 382 ($214^{+147}_{-63}$ keV, Ballantyne et al., submitted). The broad band analysis of NGC 2110 presented in Sect. \ref{380analysis} leads to a lower limit on the high-energy cutoff, $E_{\rm c}> 210$ keV. 
The unabsorbed 2--10 keV luminosity ranges between $L_X=$0.4--3.5$\times 10^{43}$ erg s$^{-1}$ (considering the 2003 {\it XMM} and 2012 {\it NuSTAR} observations as low and high flux states, respectively). If we use the 2--10 keV bolometric corrections presented in \citet{mar04}, the bolometric luminosity ranges between $L_{\rm bol}=$0.6--9$\times 10^{44}$ erg s$^{-1}$. A black hole mass of $M_{\rm BH}\simeq2\times 10^8\ M_{\odot}$ \citep{merl03, mbe07} leads to an Eddington luminosity $L_{\rm Edd}\simeq2.4\times10^{46}$ erg s$^{-1}$. Therefore, for NGC 2110 we estimate  $L_{\rm bol}/L_{\rm Edd}\simeq0.25$--$3.7\times10^{-2}$. This value is at the lower end of the distribution recently measured in CAIXA (Catalog of AGN in the XMM-Newton Archive) for a sample of 156 AGN \citep{bianchi09}. 

NGC 2110 is the second Seyfert galaxy kwown to unambiguously lack reprocessed emission from distant Compton-thick material and to show an iron K$\alpha$ emission line likely produced by Compton-thin material. NGC 7213 is the other source known to have similar properties \citep{bianchi03b, bianchi08b, lob10}. The latter is also accreting at a low Eddington rate ($L_{\rm bol}/L_{\rm Edd}\sim3\times10^{-3}$) and \citet{bianchi08b} reported a Broad Line Region origin for the iron K$\alpha$ line. Interestingly, even though the two sources show remarkably similar spectral features in the X-rays, a high energy cutoff $E_{\rm c}=95^{+50}_{-20}$ keV was detected in NGC 7213 in the simultaneous {\it XMM}-{\it BeppoSAX} data \citep{bianchi04}. 

In the near future, more AGN will be observed by {\it NuSTAR} and we will be able to investigate the coronal properties (temperature, geometry, link to the reflected emission from the accretion disk) with greater detail in additional low accretion rate objects.

\subsection{Iron K$\alpha$ temporal behavior}
We analyzed the 3-10 keV spectra of NGC 2110 from several X-ray observatories, spanning a period of 16 years. Large variations in the observed flux of the source and in the Fe K$\alpha$ line flux and equivalent width are apparent.
If the Fe K$\alpha$ line is produced by cold, distant matter we expect a constant line flux and an EW linearly anti-correlated with the intrinsic flux of the source. On the other hand, if the Fe K$\alpha$ line-emitting material is closer than distances corresponding to the time scales between the observations we expect a constant EW and a line flux linearly correlated with the illuminating continuum. We instead find a correlation between the Fe K$\alpha$ flux and the 3-10 keV observed flux $ F_{K\alpha}\propto 0.42\times(F\rm^c_{3-10})$ and an anti-correlation between the Fe K$\alpha$ EW and the 3-10 keV observed flux $EW\propto-8.2\times (F\rm^c_{3-10})$. 
We therefore propose a scenario where the Fe K$\alpha$ line is the sum of two distinct components, one constant and produced from material distant from the nucleus (the putative ``torus'') and the other one variable and linearly correlated with the primary flux that is likely associated with the Broad Line Region. In this scenario the intercept of the relation between the Fe K$\alpha$ flux and the 3-10 keV observed flux gives the amplitude of the constant component. We find a value of $(5.2\pm1.0)\times 10^{-5}$ ph cm$^{-2}$~s$^{-1}$, perfectly consistent with the resolved {\it Chandra} component (Table \ref{refl_best_fit}). The Compton reflection fraction associated to this constant iron K$\alpha$ component is consistent with the one found in the {\it NuSTAR} + {\it Swift} analysis (Sect. \ref{380analysis_swift}).

The emitting structure responsible for the variable component of the Fe K$\alpha$ line, closer to the nucleus than the constant one, was  discussed in \citet{elt07}. These authors detected a modest broadening of the Fe K$\alpha$ line in multi-epoch coadded HEG spectra, with a line width $\sigma=4500^{+3000}_{-2200}$ km s$^{-1}$. We speculate that this structure could be the same responsible for the broad, double-peaked H$\alpha$ lines (FWHM $\sim13,000$--$17,000$ km s$^{-1}$) detected in optical spectro-polarimetric observations of this source \citep{mbe07,tran10}.

 The {\it Astro-H} satellite, with its unprecedented combination of spectral resolution and collecting area at 6-7 keV, will allow us to resolve the iron line profile at 6.4 keV in NGC 2110 and determine the location of the circumnuclear emitting material and its Fe abundance.

\subsection{A closer look to the circumnuclear environment}
It is interesting to note that NGC 2110 appears to show features at infrared wavelenghts arising from complex circumnuclear regions. If we consider the mid-infrared luminosity $L_{\rm12 \mu m}=1.0\times 10^{43}$ erg/s, reported in \citet{hkg10} and the 2--10 keV luminosity range discussed above, NGC 2110 lies on the $L_{\rm MIR}\propto (L_X)^{1.11}$ relation inferred in \citet{ghs09}, who analized a sample of 42 Seyfert galaxies. This indicates that the geometry of the infrared emitter in NGC 2110 does not differ dramatically from other Seyfert galaxies. To produce enough IR continuum emission, the main requirement is enough dust with a sufficient covering factor to reprocess the intrinsic AGN power. This could be satisfied by a number of models, including a Compton-thin torus, or an extended dusty wind \citep[][and references therein]{hkt13}. In this way, one could potentially have strong dust emission, with very little accompanying X-ray Compton reflection. 

In addition, NGC 2110 is one of the few Seyfert galaxies (together with NGC 7213) that also shows Silicate dust features in emission \citep{hkg10}. The origin of Silicate emission is not fully understood even in Type 1 AGN, with non-standard dust grain properties, emission from a clumpy torus with a relatively small number of dust clumps, or emission from an extended dusty component  in the Narrow Line Region, all invoked as possible sources \citep{ssl05, srh06}. \citet{mls09} discussed in detail such a feature and found that the mid-IR component in NGC 2110 cannot be extended more than 32 pc, ruling out extended reflecting clouds as seen in NGC 4945 \citep{mrw12}. We conclude that a standard dusty torus/dusty wind model, but with a gas column density in the Compton-thin regime 
could explain the X-ray and mid-IR characteristics of NGC 2110. 

\citet{bdc10} reconstructed the spectral energy distribution of NGC 2110 using simultaneous {\it INTEGRAL} and {\it Swift} data taken in 2008 and 2009, reporting features usually shown by radio-loud sources. However, we do not confirm the flat photon index and cutoff energies they report.
We use the radio fluxes at 6 cm \citep[$F_{\rm 6cm}\simeq165$ mJy,][]{gwb95} and 20 cm \citep[$F_{\rm 20cm}\simeq300$ mJy,][]{bjf11} and compare them with the 2-10 keV fluxes we found. No significant deviations from relations between radio and X-ray emission usually found in radio-quiet Seyferts are present \citep[][and references therein]{panessa07,bianchi09b}. \citet{elk06} analyzed the {\it Chandra}, {\it HST}, and {\it VLA} imaging observations and found a small radio jet (extended by $\sim5''$ accross the nucleus). However, the authors discarded the possibility of a synchrotron origin for the X-ray emission in NGC 2110, since the radio and X-ray emission are not spatially coincident.

\section{Summary and conclusions}
We report a multi-epoch X-ray spectral analysis of the bright Seyfert 2 galaxy NGC 2110, spanning a period of 16 years. We focus on recent observations of the source with {\it NuSTAR} in 2012, when the source was at the highest flux level ever observed, and in 2013, when the source had more typical flux levels.  

Our results can be summarized as follows:
\begin{itemize}
\item a high energy cutoff $E_{\rm c}> 210$ keV has been inferred, with an upper limit on the Compton reflection contribution of $R<0.14$, confirming results from past high-energy {\it BeppoSAX} and {\it Suzaku} observations \citep{mbc99,rfk06,rmr14};
\item when multi-epoch data are considered,we find a correlation between Fe K$\alpha$ EW and intrinsic 3-10 keV flux ($EW\propto -8.2\times F\rm^c_{3-10}$) and an anti-correlation between Fe K$\alpha$ flux and intrinsic 3-10 keV flux ($ F_{K\alpha}\propto0.42\times F\rm^c_{3-10}$);
\item the Fe K$\alpha$ line is likely the sum of two components: one constant (originating from distant Compton-thick material) and the other one variable and linearly correlated with the source flux (from matter at distances compatible with the Broad Line Region). Using \textsc{MyTorus} self-consistent modeling, we find that the line could be produced by scattering material with a global covering factor of 0.5 with $N_{\rm H}=2.0\pm1.1 \times10^{23}$ cm$^{-2}$;
\item the source presents remarkably similar features to the low accretion rate Seyfert 1 galaxy NGC 7213 in the X-ray band (lack of Compton reflection, contribution from Compton-thin material to the Fe K$\alpha$ line emission) and in the infrared, where Silicate dust emission was reported in \citet{hkg10}.
\end{itemize}

\section*{ACKNOWLEDGEMENTS}
AM and GM acknowledge financial support from Italian
Space Agency under grant ASI/INAF I/037/12/0-011/13 and
from the European Union Seventh Framework Programme
(FP7/2007-2013) under grant agreement n.312789. MB
acknowledges support from the International Fulbright Science and
Technology Award. This work was supported under NASA Contract No. NNG08FD60C, and
made use of data from the {\it NuSTAR} mission, a project led by
the California Institute of Technology, managed by the Jet Propulsion
Laboratory, and funded by the National Aeronautics and Space
Administration. We thank the {\it NuSTAR} Operations, Software and
Calibration teams for support with the execution and analysis of
these observations.  This research has made use of the {\it NuSTAR}
Data Analysis Software (NuSTARDAS) jointly developed by the ASI
Science Data Center (ASDC, Italy) and the California Institute of
Technology (USA).
\bibliographystyle{mn2e}
\bibliography{sbs} 

\end{document}